\documentclass[pss]{wiley2sp}
\usepackage{amsmath,amssymb}
\usepackage{bm}

\newcommand{\abs}[1]{\ensuremath{|#1|}}

\newcommand{\braket}[2]{\ensuremath{\langle #1|#2\rangle}}

\newcommand{\braketop}[3]{\ensuremath{\langle #1|#2|#3\rangle}}

\newcommand{\bra}[1]{\ensuremath{\langle #1|}}
\newcommand{\ket}[1]{\ensuremath{|#1\rangle}}
\newcommand{\ketbra}[2]{\ensuremath{\ket{#1}\!\bra{#2}}}

\newcommand{\sgn}{\ensuremath{\operatorname{sgn}}}
\newcommand{\up}{{\uparrow}}
\newcommand{\down}{{\downarrow}}

\newcommand{\void}[1]{}

\begin{document}
	\title{Entanglement-symmetry control in a quantum-dot Cooper-pair splitter}
	\titlerunning{Entanglement-symmetry control }
	
	\author{%
	Robert Hussein\textsuperscript{\textsf{\bfseries 1,2}},
	Alessandro Braggio\textsuperscript{\textsf{\bfseries 1,3,4}},
	Michele Governale\textsuperscript{\Ast,\textsf{\bfseries 5}}
	}
	
	\authorrunning{Robert Hussein et al.}
	
	\mail{e-mail \textsf{Michele.Governale@vuw.ac.nz}%
	}
	
	\institute{%
	\textsuperscript{1}\,SPIN-CNR, Via Dodecaneso 33, 16146 Genova, Italy\\
	\textsuperscript{2}\,Fachbereich Physik, Universit\"at Konstanz, D-78457 Konstanz, Germany\\
	\textsuperscript{3}\,NEST, Istituto di Nanoscienze CNR-NANO, Piazza San Silvestro 12, 56127, Pisa, Italy\\
	\textsuperscript{4}\,INFN, Sez. Genova, Via Dodecaneso 33, 16146 Genova, Italy\\
	\textsuperscript{5}\,School of Chemical and Physical Sciences and MacDiarmid Institute for Advanced Materials and Nanotechnology, 
	Victoria University of Wellington, P.O. Box 600, Wellington 6140, New Zealand
	}
	
	\received{XXXX, revised XXXX, accepted XXXX} %
	\published{XXXX} %
	
	
	\abstract{%
	\abstcol{The control of nonlocal entanglement in solid state systems is a crucial ingredient of quantum technologies. We investigate a Cooper-pair splitter  based on a double quantum dot realised in a semiconducting nanowire. In the presence of interdot tunnelling the system provides a simple mechanism to develop spatial entanglement even in absence of nonlocal coupling with the superconducting lead.  We discuss the possibility to control the symmetry (singlet or triplet) of spatially separated, entangled electron pairs taking advantage of the spin-orbit coupling of the nanowire.  We also demonstrate that the spin-orbit coupling does not impact over the entanglement purity of the nonlocal state generated in the double quantum dot system.
	}{}
	}
	
	\titlefigure[height=3.1cm]{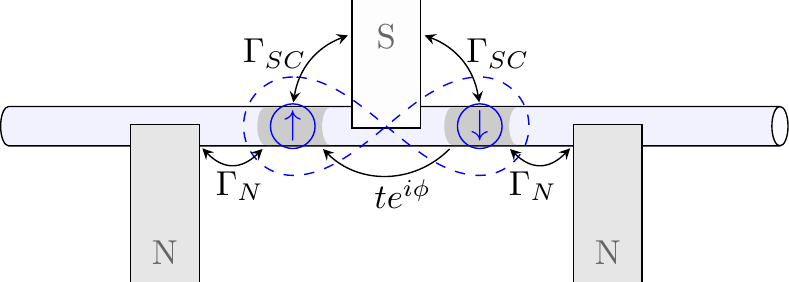}
	\titlefigurecaption{
	Cooper-pair splitter realised in a nanowire acting as double quantum dot which is coupled to 
	one superconductor (S) and two normal leads (N). Electron singlets in the superconductor
	are split nonlocally into the different dots as schematically shown in the figure. The entangled electrons in the two quantum dots are then transferred by tunnelling, with rate $\Gamma_N$,  to the normal leads.}

	\maketitle 
	
	\section{Introduction}
	
	The generation and control of entangled electron pairs in solid-state systems is a key technology for quantum computation and quantum 
	information \cite{MonroeNature2002a,LaddNature2010a}. In particular, semiconductor--superconductor hybrid devices 
	connected to normal metal leads are seen as promising 
	candidates \cite{LesovikEPJB2001a,RussoPRL2005a,HofstetterNature2009a,HerrmannPRL2010a,Martin-RoderoAP2011a,RoddaroNanoR2011a,SchindelePRL2012a,RomeoNanoL2012a,BrauneckerPRL2013a,SchindelePRB2014a,RossellaNatNano2014a} for entanglement
	generation and for the emerging field of superconducting spintronics \cite{LinderNatPhys2015a}. 
	The electrons forming a Cooper pair in a $s$-wave superconductor are in an entangled state and by spatially splitting them into different normal leads, one can inject  entangled pairs in the leads.
	This nonlocal Cooper-pair transfer process, also 
	known as cross-Andreev reflection (CAR), is the cornerstone for the production of 
	two spatially separated entangled electrons. 
	This process is in competition with local Andreev reflection (LAR), where the two electrons in a Cooper pair enter the same lead. Obviously, LAR does not directly contribute to the nonlocal 
	entanglement generation. An experimentally pragmatic solution is 
	to couple the superconductor and the normal leads to a double-quantum dot (DQD)  where large intradot Coulomb repulsion makes LAR energetically unfavourable 
	\cite{ChoiPRB2000a,RecherPRB2001a,SauretPRB2004a,HofstetterNature2009a,HerrmannPRL2010a,SchindelePRB2014a,FueloepPRL2015a}.
	Finally, recent theoretical investigations on nanowires in contact with two conventional $s$-wave superconductors predict that strong spin-orbit (SO) interaction may induce triplet superconducting  correlations in the nanowire   \cite{ShekhterPRL2016a,YuPRB2016a}. Unconventional superconducting correlations can be generated in double quantum dots by means of nonlocal magnetic fields \cite{SothmannPRB2014a}. 
	Other studies considered the possibility to use a spin-selective mechanism, such as ferromagnetic leads, in order to quantify better the entanglement generated in the DQD Cooper pair splitter  \cite{BeckmannPRL2004a,HofstetterPRL2010a,SammJAP2014a,KlobusPRB2014a,TrochaPRB2015a}.
	
	We have shown in Ref.~\cite{HusseinPRB2016a} that for finite Coulomb interaction, in the presence of interdot tunnelling and strong local coupling with the superconductor, the Cooper-pair splitter may develop nonlocal entanglement even without an explicit nonlocal coupling. This is the result of a coherent process where a Cooper-pair is transferred to one quantum dot leading to a virtually doubly occupied state and the subsequent transfer of one of the 
	electrons by interdot tunnelling to the other quantum dot. In other words, the entangled electron pair in one dot  is coherently converted into a nonlocally  entangled pair where each electron resides in a different dot. 
	In this paper  we investigate how to control the symmetry of the entangled states in the DQD. In particular, we show that, in the presence of a SO contribution in the interdot tunnelling, one can change the symmetry of the entanglement not only from singlet to triplet \cite{HusseinPRB2016a} but, intriguingly, also to a linear combination that still preserves the entanglement. 
	
	The paper is organised as follows: In section~\ref{sec.:model}, we briefly introduce our model and describe how to derive currents based on the
	master-equation approach; Section~\ref{sec.:CAR} is devoted to the control of the entanglement. In section~\ref{sec.:conclusions}, we summarise our main results.

	\section{\label{sec.:model}Model and formalism}
	In this section we briefly introduce the model for the  Cooper-pair splitter sketched in the figure in the abstract,
	and describe the formalism used to calculate the stationary current. 
	
	The system under study consists of a double quantum dot  tunnel coupled to two normal metal leads 
	and to one $s$-wave superconductor. The DQD system is described by the Hamiltonian
	\begin{align}
H_{\textrm{DQD}}= { }&
	 \sum_{\alpha,\sigma
	}\varepsilon_{\alpha} n_{\alpha\sigma}
	+U_C\sum_{\alpha} n_{\alpha\uparrow}n_{\alpha\downarrow}
\nonumber\\
&+
	\frac{t}{2}\sum_{\sigma}\big( e^{i\sgn(\sigma)\phi} d^\dag_{L\sigma}d_{R\sigma} +\mathrm{H.c.}
\big) \label{eq.:HDQD}
	\end{align}
	with $\varepsilon_{\alpha}$ the orbital level of dot $\alpha=L,R$ and $U_C$ the intradot Coulomb energy. For simplicity, we assume the interdot
	Coulomb interaction to be negligible since the CAR physics discussed in this paper is not affected substantially  by the presence of this term.
	At most 2 electrons of opposite spin $\sigma=\up,\down$ can reside in one quantum dot, which leads to a maximal total
	occupation of 4 electrons in the DQD system. Here, $n_{\alpha\sigma}=d^\dag_{\alpha\sigma}d_{\alpha\sigma}$
	is the number operator with corresponding 
	fermionic creation operator $d^\dag_{\alpha\sigma}$. The last term in Eq.~\eqref{eq.:HDQD} describes the tunnelling between the dots in the presence of SO interaction. 
	We use the notation: $\sgn(\up)=1$ and $\sgn(\down)=-1$.
	The physical behaviour can be described with a SO angle $\phi=k_{\textrm{SO}}l$ spanning
	from $-\pi/2\leq \phi\leq \pi/2$ and a real interdot tunnelling amplitude $t$;
	$k_{\textrm{SO}}$ encodes the SO strength and $l$ is the distance between the dots.
	
	The subgap physics for large superconducting gap can be described by an effective  Hamiltonian \cite{RozhkovPRB2000a,MengPRB2009a,EldridgePRB2010a,BraggioSSC2011a,SothmannPRB2014a},
	\begin{align}
		\begin{split}
H_{S}=H_{\textrm{DQD}} & - \frac{\Gamma_{SC}}{2}\sum_{\alpha}  \big( 
	d^\dagger_{\alpha,\uparrow} d^\dagger_{\alpha,\downarrow}+\mathrm{H.c.}
\big)\\
& -\frac{\Gamma_S}{2}\big( 
	d^\dagger_{R,\uparrow} d^\dagger_{L,\downarrow}- d^\dagger_{R,\downarrow} d^\dagger_{L,\uparrow}+ \mathrm{H.c.}
\big),  \label{eq.:Heff}
		\end{split}
	\end{align}
	where the second term describes the local tunnelling process of a Cooper-pair into one dot with tunnelling rate $\Gamma_{SC}$ and 
	the last term describes the nonlocal tunnelling process of a Cooper-pair splitting into both dots with the rate $\Gamma_{S}\sim\Gamma_{SC}e^{-l/\xi}$ 
	which is maximal when the distance between the dots is much smaller than the superconducting coherence length $\xi$ \cite{RecherPRB2001a}.
	\begin{figure*}[ht]
		\includegraphics[width=\textwidth]{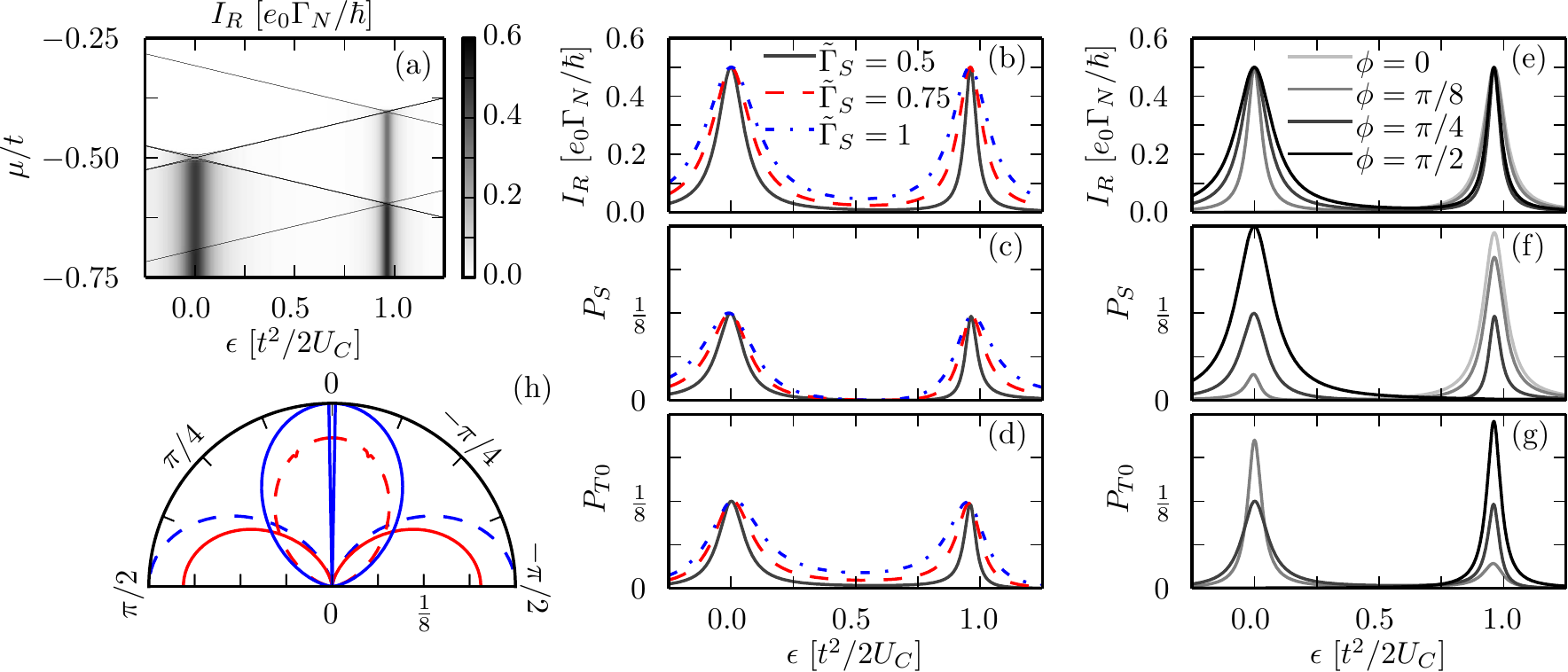}\caption{\label{fig.:CAROverview}%
		(a) Current $I_R$ as function of the level position $\varepsilon=\varepsilon_L=\varepsilon_R$ and the chemical potential $\mu=\mu_L=\mu_R$
		for the local tunnel coupling $\Gamma_{SC}=2\Gamma_{S}=U_C/200=20k_BT=200\Gamma_N$, tunnelling amplitude $t=0.2 U_C$, and the 
		spin-orbit angle $\phi=\pi/4$. The solid lines indicate the Andreev spectrum. 
		(b,e) Current $I_R$ at constant $\mu=-t$ as function of the level position for various values of the nonlocal coupling $\tilde\Gamma_{S}=\Gamma_{S}/\Gamma_{SC}$ 
		[panel (b)] and SO angle $\phi$ [panel (e)]. (c,d) Singlet population $P_S$ and unpolarised triplet population 
		$P_{T0}$ corresponding to the current in panel (b). (f,g) Singlet population $P_S$ and unpolarised triplet population 
		$P_{T0}$ corresponding to the current in panel (e). 
		(h) Polar plot of the singlet population $P_S$ (dashed lines) and unpolarised triplet population $P_{T0}$ (solid lines) as a function of the SO angle $\phi$
		for the non-shifting peak (blue lines) and the shifted (red lines). The polar plot has a radius of 1/4 that corresponds to the maximal occupation probability [see also panel (f) and (g)].
		}
	\end{figure*}
	
	The Hamiltonian $H_{\text{S}}$ can be written in diagonal form as $H_{S}=\sum_a E_a\ketbra{a}{a}$. It is easy to show that in the basis of the eigenstates $\ket{a}$, the dynamics of the off-diagonal elements of the reduced density matrix of the proximised DQD is decoupled from the occupation probabilities and it does not enter in the expressions for the currents. Therefore we can restrict to the occupation probabilities $P_a$  for the eigenstate  $\ket{a}$ of  $H_{S}$,   which obey the master equation  $\dot P_a = \sum_{a'}\big( w_{a\leftarrow a'}P_{a'}- w_{a'\leftarrow a}P_{a}\big)$ in lowest order in the tunnelling rates to the normal leads \cite{BagretsPRB2003a,BraggioPRL2006a,KaiserAP2007a,FlindtPRL2008a,FlindtPRB2010a,HusseinPRB2014a}. 
	The tunnelling rates $w_{a\leftarrow a'}$ for the transition from the state $\ket{a'}$ to the state $\ket{a}$ can be obtained by  Fermi's golden rule  \cite{BraggioSSC2011a,HusseinPRB2016a}
	\begin{align}
	w_{a\leftarrow a'}(\bm\chi)
	{=} \Gamma_{N}&\sum_{\alpha\sigma}e^{i\chi_\alpha} \big[1-f_\alpha(E_{a'}-E_a)\big]\abs{\braketop{a}{d_{\alpha\sigma}}{a'}}^2\nonumber\\
	    +\Gamma_{N}&\sum_{\alpha\sigma}e^{-i\chi_\alpha} f_\alpha(E_a-E_{a'})\abs{\braketop{a}{d_{\alpha\sigma}^\dag}{a'}}^2
	   \label{eq.:w}
	\end{align}
	with $f_\alpha(\varepsilon)=\{1+\exp[(\varepsilon-\mu_\alpha)/k_BT]\}^{-1}$ being the Fermi function of the normal lead $\alpha$ with chemical 
	potential $\mu_\alpha$ and temperature $T$. 
	We set the chemical potential of the superconducting lead to zero.  
	The tunnelling rates with the normal leads $\Gamma_{N}$ are assumed to be energy independent in the energy window relevant for the transport.
	The counting field \cite{BagretsPRB2003a,FlindtPRL2008a,FlindtPRB2010a} ${\bm\chi}=(\chi_L,\chi_R)$ in Eq.~\eqref{eq.:w} allows us
	to express the stationary current through the right normal lead via
	\begin{align}
\label{eq:cur}
	I_R = -i\frac{e_0}{\hbar}\sum_{a,a'}\frac{\partial w_{a\leftarrow a'}(\bm\chi)}{\partial\chi_R}\Big|_{{\bm\chi}={\bm0}}P_{a'}^{\textrm{stat}},
	\end{align}
	where $P_{a'}^{\textrm{stat}}$ is the stationary solution of the master-equation and $e_0$ denotes the
	electron charge. We finally note that hereafter we consider the temperature being $\Gamma_{SC} \gg k_BT\gg \Gamma_N$ 
	where the first inequality guarantees highly resolved Andreev states and the second inequality 
	guarantees the validity of the first-order perturbation theory in $\Gamma_N$.

	\section{\label{sec.:CAR}Resonant current peaks and entanglement}
	In the following, we will focus on the CAR resonance of the system where the entanglement 
	is maximal. We  assume a completely symmetrical DQD configuration. 
	The asymmetry of the coupling of the two dots with the normal and superconducting leads may be easily included in this scheme and does not crucially modify the reported results. The interested reader can find a more detailed discussion of the effect of structural asymmetry, including different Coulomb energies for the two dots, in the Appendix B of Ref.~\cite{HusseinPRB2016a}.
	We consider symmetric bias voltages 
	$\mu\equiv\mu_L=\mu_R$ and level positions $\varepsilon\equiv\varepsilon_L=\varepsilon_R$ which is the optimal point to develop spatial entanglement in the system. 
	
	The nonlocal coupling to the superconductor, for sufficiently negative chemical 
	potential $\mu$, leads to nonlocal Cooper-pair splitting with a resonant current at zero dot level $\varepsilon\approx0$, as shown in Fig.~\ref{fig.:CAROverview}(a). Spin-independent interdot tunnelling may shift the position at which this resonant current occurs to positive level positions $\varepsilon>0$ for $\phi=0$. In general, an additional resonance emerges in the presence of
	SO interaction if $\phi\neq k\pi$ with $k$ integer [see Fig.~\ref{fig.:CAROverview}(a)]. 
	In Ref. \cite{HusseinPRB2016a} we discussed only the two cases $\phi=0$ and $\phi=\pi/2$ for the SO angle, which are representative 
	for the behaviour of the system with interdot tunnelling when the SO effect is absent or when it is maximal, respectively. These two cases crucially differ in the symmetry of the entanglement developed in the DQD at the nonlocal resonance. 
	Intriguingly for $\phi=\pi/2$ a nonlocal triplet is generated even if the $s$-wave superconducting lead injects in the DQD only singlet Cooper-pairs. This symmetry change is brought about by the SO interaction.
	In the following section, we will discuss the rich phenomenology determined by a generic SO angle $\phi$.
	\subsection{Effect of SO angle $\phi$ over the entanglement symmetry}
	Figure~\ref{fig.:CAROverview}(a) shows the current through the right lead $I_R$ as a function 
	of the dot level $\varepsilon$ and the chemical potential $\mu$. 
	For finite interdot tunnelling amplitude $t$, with $0< \Gamma_S,\Gamma_{SC}\ll t\ll U_C$, and for intermediate SO angle $\phi=\pi/4$, the CAR peak is in general split in
	two resonances at $\varepsilon\approx0$ and $\varepsilon\approx t^2/2U_C$ where the one at zero level position sets in
	for $\mu\lesssim-t/2$. This is very similar to the phenomenology discussed for the case $\phi=\pi/2$.\footnote{A notable difference is that for a generic angle $|\phi|\neq\pi/2$ 
	the current shows a particle-hole (PH) symmetry breaking behaviour of the current in the plane $(\varepsilon,\mu)$ when the nonlocal coupling $\Gamma_S\neq0$, see detailed discussion on PH symmetry breaking in Ref.~\cite{HusseinPRB2016a}.}

	In Fig.~\ref{fig.:CAROverview}(b) we show the current as a function of the level position $\varepsilon$ at a fixed value of the chemical potential $\mu=-t$
	for different values of the the nonlocal tunnel coupling $\Gamma_S$ (different line styles). The dependence on $\Gamma_S$ is very weak and for this reason one can keep the nonlocal coupling  fixed at an intermediate value $\Gamma_S=\Gamma_{SC}/2$ without loosing much generality.
	
	In order to discuss the symmetry of the nonlocal entanglements in the DQD, we show in panels (c) and (d) the populations $P_S$ and $P_{T0}$, respectively, for the singlet state,
	$\ket{S}=\frac{1}{\sqrt{2}} \big( d^\dag_{R\uparrow} d^\dag_{L\downarrow} -d^\dag_{R\downarrow} d^\dag_{L\uparrow}\big)\ket{0}$,
	and the unpolarised triplet state, 
	$\ket{T0}=\frac{1}{\sqrt{2}}\big( d^\dag_{R\uparrow} d^\dag_{L\downarrow}+d^\dag_{R\downarrow} d^\dag_{L\uparrow}\big)\ket{0}$. 
	The populations at the maximum of the current peaks are a very rough estimate of the entanglement developed in the system. We first notice that for $\phi=\pi/4$ both CAR peaks are characterised by finite  singlet and triplet populations---the maximum value of the populations is essentially independent of $\Gamma_S$.  
	The effect of the SO angle $\phi$ on the populations of the DQD is clearly shown in the panels (e)-(g). 
	The shifted peak (the right one) at $\varepsilon\approx t^2/2U_C$ is for $\phi=0$ essentially characterised by the singlet symmetry. For higher values of $\phi$ the singlet population is reduced and an almost equivalent triplet component emerges, and for maximal SO angle $|\phi|=\pi/2$ only the triplet component remains. 
	The non-shifting peak $\varepsilon\approx0$, however, exhibits an almost opposite behaviour: for small $\phi$, it has essentially triplet symmetry 
	and for increasing SO angle this unpolarised-triplet population is progressively transferred to the singlet population which becomes maximal for $\phi=\pi/2$. 
	The SO-angle $\phi$ turns out to be an effective control knob to modify the proportion of the singlet/triplet populations at the  current resonances as can be appreciated in panel (h) of Fig.~\ref{fig.:CAROverview}, which shows the singlet population $P_S$ (dashed lines) and the unpolarised-triplet population $P_{T0}$ (solid lines) as a function of $\phi$ at the non-shifting (blue) and the shifted (red) current-peak maxima.
	
	The value of the nonlocal coupling $\Gamma_S$ does not change the position of the current peaks, but affects the linewidth of 
	both resonances. We have already shown in Ref.~\cite{HusseinPRB2016a} that entanglement in the system may still be generated even without nonlocal coupling $\Gamma_S=0$. However, in the following we do not specifically consider this case, since to obtain the maximal nonlocal  entanglement  it is advantageous to keep the nonlocal coupling finite. 
	In the next section, we introduce a minimal model which describes how the SO angle controls the symmetry of the entanglement in the DQD. 
	
	\begin{table*}[ht]
		\begin{tabular}{@{}lllll@{}}
			\hline
			$\epsilon_{0}$ & $=0$ & & $\ket{0}$ & \\
			$\epsilon_{1}$ & $=2\varepsilon$ & & $\ket{\epsilon_1}$ & $=i\sin(\phi)\ket{S}+\cos(\phi)\ket{T0}$ \\
			$\epsilon_{2}$ & $=2\varepsilon+U_C$ & & $\ket{\epsilon_{2}}$ & $=\ket{d-}$ \\
			$\epsilon_\pm$ & $=2\varepsilon+\frac{1\pm\sqrt{1 + 4\tau^2}}{2}U_C$ & &
			$\ket{\epsilon_\pm}$ & $=\sgn(\tau)\sqrt{\frac{1\mp\alpha}{2}}\big(\cos(\phi)\ket{S}+i\sin(\phi)\ket{T0}\big)\pm\sqrt{\frac{1\pm\alpha}{2}}\ket{d+}$\\
			\hline
		\end{tabular}
		\caption{\label{tab.:HS_spec} Eigenvalues and eigenvectors of the Hamiltonian $H_0$, Eq.~\eqref{eq.:H0}, in the 
		limit $\Gamma=\Gamma_{SC}=0$ with $\alpha=1/\sqrt{1+4\tau^2}$, $\ket{d\pm}=(\ket{dR}\pm\ket{dL})/\sqrt{2}$  and $\tau=t/U_C$. 
		}
	\end{table*}
	\begin{figure}[t]
		\includegraphics{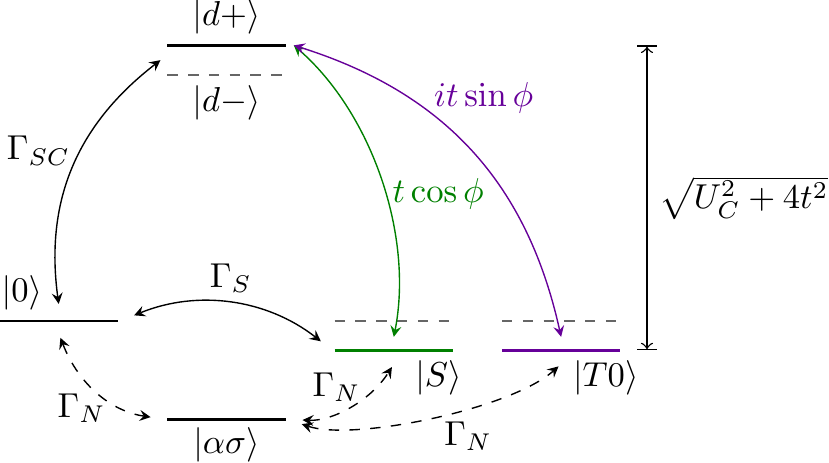}\caption{\label{fig.:eff_level_struc}%
		Effective level structure at the CAR resonances. 
		}
	\end{figure}
	\subsection{\label{sec.:spectrum}Level structure at the nonlocal resonances}
	In Fig.~\ref{fig.:eff_level_struc} we show the minimal model which  
	describes the occurrence of the nonlocal current resonances shown in Fig.~\ref{fig.:CAROverview}.
	It is essentially an extension of the model introduced in Ref.~\cite{HusseinPRB2016a} to a generic SO angle and it is able to capture the essential physics of the SO control over the entanglement symmetry in the DQD.  This model contains only the empty state $\ket{0}$, the single occupancies $\ket{\alpha\sigma}$ of the dot $\alpha=L,R$ with spin $\sigma=\uparrow,\downarrow$, the singlet state $\ket{S}$, the unpolarised triplet state $\ket{T0}$, and the doubly occupied states
	$\ket{d\alpha}= d_{\alpha\uparrow}^\dag  d_{\alpha\downarrow}^\dag \ket{0}$.\footnote{We do not include the polarised triplet states $\ket{T\sigma}=d_{R\sigma}^\dag  d_{L\sigma}^\dag \ket{0}$ since they are fully decoupled from the other even-parity states.} The arrows correspond to the different coupling mechanisms between these states.
	We can limit our investigation of the spectrum and eigenstates to the even-parity sector since $\Gamma_N\ll \Gamma_{SC},\Gamma_S,t$ 
	and the tunnelling rates with the normal lead $\Gamma_N$ correspond to transitions between even-sector states and the singly occupied states $\ket{\alpha\sigma}$.
	In the even sector, chosing the basis $\{\ket{0},\ket{S},\ket{T0},\ket{dL},\ket{dR}\}$, the Hamiltonian reads
	\begin{align}
			H_0 = \begin{pmatrix}
	       0 && -\frac{\Gamma_S}{\sqrt{2}}&& 0 && -\frac{\Gamma_{SC}}{2} && -\frac{\Gamma_{SC}}{2} \\
	       -\frac{\Gamma_S}{\sqrt{2}} && 2\varepsilon && 0 && \frac{t\cos\phi}{\sqrt{2}} && \frac{t\cos\phi}{\sqrt{2}}\\
	       0 && 0 && 2\varepsilon && \frac{it\sin\phi}{\sqrt{2}} && \frac{it\sin\phi}{\sqrt{2}}\\
	       -\frac{\Gamma_{SC}}{2} && \frac{t\cos\phi}{\sqrt{2}} && -\frac{it\sin\phi}{\sqrt{2}} && 2\varepsilon+U_C && 0\\
	       -\frac{\Gamma_{SC}}{2} && \frac{t\cos\phi}{\sqrt{2}} && -\frac{it\sin\phi}{\sqrt{2}} && 0 && 2\varepsilon+U_C\\
			      \end{pmatrix}.
	\label{eq.:H0}
	\end{align}
	We note the SO angle dependence of the interdot tunnelling terms in the off-diagonal elements of $H_0$, which is essentially 
	determined by the different symmetry of the singlet and the triplet state. 
	Since $\Gamma_{S},\Gamma_{SC}\ll t$, we can take the  limit $\Gamma_S=\Gamma_{SC}=0$ and the eigenvalues and eigenvector in this case are given in Table~\ref{tab.:HS_spec}. 
	There, we have introduced the symmetric and anti-symmetric combinations of the doubly-occupied states $\ket{d\pm}=(\ket{dR}\pm\ket{dL})/\sqrt{2}$ which arise from the breaking of the degeneracy of the states $\ket{d\alpha}$  by the interdot tunnelling. We also introduced the dimensionless variable $\tau=t/U_C$.\footnote{In this work we consider the intradot Coulomb energies of the two dots to be identical but essentially the same physics occurs if they differ by less than $\Gamma_{SC}(t/U_C)$.} 
	
	In Table~\ref{tab.:HS_spec} one immediately sees that the eigenvalues of $H_0$ are independent of the SO angle $\phi$. 
	This is in full agreement with the fact that the peak position for the non-shifting (shifted) current resonance in Fig.~\ref{fig.:CAROverview} is related to $\ket{\epsilon_1}$ ($\ket{\epsilon_-}$) and is independent (dependent) of $\tau$ and independent of $\phi$.\footnote{The position $\varepsilon_{-}=\big(\sqrt{1+4\tau^2}-1\big)U_C/4$ of the shifted current resonance as determined by requiring the degeneracy of $|\epsilon_-\rangle$ with $|0\rangle$, i.e. $\epsilon_-(\varepsilon_-)=0$.} This fully explains the peculiarity that only one
	of the resonance peaks shifts with increasing $\tau$ independent from $\phi$.
	The coupling strength $\tau$ and the SO angle $\phi$ enter in the expression of the eigenstates $\ket{\epsilon_\pm}$, which is a result of the peculiar level repulsion between the state $\ket{d+}$ and the singlet and triplet degenerate space.  
	
	\section{Nonlocal entanglement control}
	In this section we discuss the role of the SO angle $\phi$ as a control parameter to modify the symmetry of the 
	nonlocal entanglement in the DQD system. 
	
	The possibility to control the symmetry of the entanglement is suggested by the aforementioned mutual and opposite symmetry character of the DQD populations at the current peaks discussed in Fig.~\ref{fig.:CAROverview}(h). This result can easily understood by inspecting the angle dependence of the eigenstate decomposition reported in Table~\ref{tab.:HS_spec} of the eigenstate $\ket{\epsilon_1}$ ($\ket{\epsilon_-}$) for the left (right shifted)  peak in terms of the singlet and triplet components . 
	In the Cooper pair splitter one may assume that the entanglement symmetry of the state with one electron residing on each dot [i.e the (1,1) charge state] can be transferred to the leads. We need to verify that the linear superposition of singlet and triplet realised in the DQD is still entangled for all values of the angle $\phi$.
	
	In the literature it has been discussed \cite{KlobusPRB2014a,TrochaPRB2015a} how to employ ferromagnetic leads to evaluate the entanglement and other quantum properties of  DQD Cooper pair splitters. 
	Modelling how the entanglement is transferred from the DQD to the leads is beyond the scope of the present work and we limit ourself to discuss the the symmetry of the entanglement of the DQD in the (1,1) charge state and how it can be controlled by the SO angle. This is an upper bound for the entanglement injected in the normal leads.
	In our case this upper bound is naturally identified with the entanglement of the DQD states which are involved in the nonlocal transport in the charge sector (1,1). We define the projection operator on that charge sector (1,1) as $\mathcal{P}_{(1,1)}$. The projection of the state relevant for the non-local transport resonance $\varepsilon\approx0$ ($\varepsilon\approx t^2/2U_C$)  is $\mathcal{P}_{(1,1)}[\ket{\epsilon_1}]=\ket{\Psi_+}$ ($\mathcal{P}_{(1,1)}[\ket{\epsilon_-}]=\sgn(\tau)\ket{\Psi_-}$) with
	\begin{align}
\binom{\ket{\Psi_+(\phi)}}{\ket{\Psi_-(\phi)}}=\cos(\phi)\binom{\ket{T0}}{\ket{S}}+i\sin(\phi)\binom{\ket{S}}{\ket{T0}}
	\end{align}
	where $\Psi_+(\phi) \equiv\ket{\epsilon_1}$ since $\ket{\epsilon_1}$ lives only in this charge sector.
	These projections are linear combination of the singlet and the unpolarized triplet states and, in principle, can be \emph{not entangled}. 
	For example if the DQD is in the state 
	$\ket{\up\down}= d^\dag_{R\uparrow} d^\dag_{L\downarrow}\ket{0}=(\ket{S}+\ket{T0})/\sqrt{2}$ is purely local in nature. 
	
	In the charge sector (1,1) the state of the DQD can be mapped into a two qubit state and we can use the concept of the concurrence to evaluate the entanglement\cite{HillPRL1997a,WoottersPRL1998a,EckertAP2002a}. The concurrence of a pure state $\ket{\psi}$ is  defined as $C(\ket{\psi})=\abs{\sum_k\braket{B_k}{\psi}^2}$ and it  ranges from zero (not entangled) to one (maximal entangled). In the DQD charge sector (1,1), the state  $\ket{B_k}$ is the $k$th element of 
	$\big\{\ket{S},i\ket{T0},(\ket{T\up}+\ket{T\down})/\sqrt{2},i(\ket{T\up}-\ket{T\down})/\sqrt{2}\big\}$ where
	$\ket{T\sigma}= d_{R\sigma}^\dag  d_{L\sigma}^\dag \ket{0}$ with $\sigma=\up,\down$ denotes the polarised triplet states. One easily finds that 
	\begin{align}
C(\ket{\Psi_\pm(\phi)})=|\cos(\phi)^2+\sin(\phi)^2|=1 
	\end{align}
	independent of the SO angle, i.e. the two projected states $\Psi_\pm(\phi)$ are fully entangled. This result means that the SO coupling modifies the symmetry of the states without destroying its nonlocality and it confirms that both resonances can develop a certain amount of nonlocal spatial entanglement. 
	
	The state corresponding to the non-shifting resonance at $\epsilon\approx0$ appears to be optimal even if, in order to activate it, we need a non vanishing nonlocal coupling $\Gamma_S$. The shifting resonance instead involves the state $\ket{\epsilon_-}$ which is mixed with the double occupancy. So we expect a degradation of the efficiency due its partial mixing with the state $\ket{d+}$ which cannot contribute to the nonlocal spin entanglement between the two dots.
	The main advantage of the shifting current resonance is that it does not require the presence of the nonlocal coupling $\Gamma_S$  and only a finite local term $\Gamma_{SC}$ is necessary. This allows to consider setups where the distance of the two dots $l\gg\xi$ with the consequent suppression of the nonlocal coupling. 
	
	We would also like to clarify the role of the nonlocal coupling $\Gamma_S$ especially for the non-shifting resonance. When $\phi=0$, the associated state $\ket{\epsilon_1}$ contains only the triplet component, see  Table~\ref{tab.:HS_spec}. In such a case the non-shifting peak completely disappears since the triplet states totally decouple from the superconducting lead and there is no way to establish a current from the superconductor to the normal leads. 
	One may wonder if the fact that the nonlocal singlet coupling $\Gamma_S$ is 
	relevant in order to observe the non-shifting resonance means that in the end the current in the leads will exhibit  only singlet symmetry. 
	This, however, is not the case for $\Gamma_S\ll t$: the nonlocal term has only the role to couple the system with the superconductor but the current symmetry is determined by the DQD Hamiltonian Eq.~\eqref{eq.:H0}. However, the fact 
	that the superconductor \emph{only} couples with the singlet component leads to a reduction of the effective coupling to $\Gamma_S|\sin(\phi)|$. This explains why at small SO angles $\phi\approx 0$ the non-shifting resonance is hardly detectable indeed, following the methods reported in Ref. \cite{HusseinPRB2016a}, we predict a linewidth $w_{0}\sim\sqrt{2} \Gamma_S|\sin(\phi)|$ for the non-shifting resonance. 
	
	Analogously, one can see that also the linewidth of the shifting resonance is dependent on the SO angle. In the limit $\tau=t/U_C\ll1$ it is given by
	\begin{equation}  
		w_{-}\sim \sqrt{2} \left| \Gamma_S\cos(\phi)-\frac{\Gamma_{SC}\tau}{\sqrt{1+4\tau^2}}\right|
	\end{equation}
	which could potentially affect the visibility of the shifted resonance for particular choices of parameters.
	
	Finally, we wish to mention that different types of processes may affect the efficiency of the transfer of the nonlocal entanglement from the DQD  to the leads and we envision further studies on this interesting point.
	 
	\section{\label{sec.:conclusions}Conclusions}
	Spatially nonlocal entanglement generation with a hybrid superconductor double-quantum dot
	Cooper-pair splitter has been investigated. We have shown that the presence of interdot tunnelling
	and SO interaction can be exploited to change the symmetry of the entanglement. In particular, the spin-orbit angle can be used as control parameter to continuously 
	change between singlet and triplet symmetry without crucially affecting the degree of entanglement. Near the current resonance at zero level position the entanglement is maximal, while at the other resonance spatially nonlocal entanglement can be generated even without nonlocal coupling with the superconductor. 

	\begin{acknowledgement}		We acknowledge discussion with S. Roddaro.
		This work has been supported by Italian's MIUR-FIRB 2012 via the HybridNanoDev project under Grant no. RBFR1236VV.
		A.B. acknowledges financial support from STM 2015, 630925-COHEAT and COST Action MP1209.
	\end{acknowledgement}
	
	\bibliographystyle{pss}
	\providecommand{\WileyBibTextsc}{}
	\let\textsc\WileyBibTextsc
	\providecommand{\othercit}{}
	\providecommand{\jr}[1]{#1}
	\providecommand{\etal}{~et~al.}

	\
\end{document}